# Monitoring transient elastic energy storage within the rotary motors of single $F_o F_1$-ATP synthase by DCO-ALEX FRET

Stefan Ernst[a,b], Monika G. Düser[a], Nawid Zarrabi[a] Michael Börsch[a,b],*

[a] 3rd Institute of Physics, University of Stuttgart, Pfaffenwaldring 57, 70550 Stuttgart, Germany
[b] Single-Molecule Microscopy Group, Jena University Hospital, Friedrich Schiller University Jena, Nonnenplan 2 - 4, 07743 Jena, Germany

**ABSTRACT**

The enzyme $F_o F_1$-ATP synthase provides the 'chemical energy currency' adenosine triphosphate (ATP) for living cells. Catalysis is driven by mechanochemical coupling of subunit rotation within the enzyme with conformational changes in the three ATP binding sites. Proton translocation through the membrane-bound $F_o$ part of ATP synthase powers a 10-step rotary motion of the ring of $c$ subunits. This rotation is transmitted to the γ and ε subunits of the $F_1$ part. Because γ and ε subunits rotate in 120° steps, we aim to unravel this symmetry mismatch by real time monitoring subunit rotation using single-molecule Förster resonance energy transfer (FRET). One fluorophore is attached specifically to the $F_1$ motor, another one to the $F_o$ motor of the liposome-reconstituted enzyme. Photophysical artifacts due to spectral fluctuations of the single fluorophores are minimized by a previously developed duty cycle-optimized alternating laser excitation scheme (DCO-ALEX). We report the detection of reversible elastic deformations between the rotor parts of $F_o$ and $F_1$ and estimate the maximum angular displacement during the load-free rotation using Monte Carlo simulations.

## 1 INTRODUCTION

ATP is known as the 'chemical energy currency' of living organisms. The membrane-bound proteins, that are continuously regenerating ATP, are called $F_o F_1$-ATP synthases. They are located in the thylakoid membrane, in the inner mitochondrial membrane and in the plasma membrane of bacteria [1] like *Escherichia coli*. The enzymes catalyze the formation of adenosine triphosphate (ATP) from adenosine diphosphate (ADP) and phosphate. $F_o F_1$-ATP synthase consists of two major parts (Figure 1). The membrane-bound $F_o$ part is coupled to the hydrophilic $F_1$ part. The $F_o$ part comprises the $a$ and the dimeric $b_2$ subunits as well as a ring of ten $c$ subunits [2]. The hydrophilic $F_1$ part consists of five different subunits with stoichiometry $\alpha_3\beta_3\gamma\delta\varepsilon$ [3]. ATP synthesis takes place in the $F_1$ part. It shows a threefold symmetry and contains three catalytic and three non-catalytic nucleotide binding sites.

Depending on ATP and ADP concentrations and a 'proton motive force', the enzyme is able to do both ATP hydrolysis and ATP synthesis. During ATP synthesis the $F_o F_1$-ATP synthase transfers electrochemical into chemical energy by mechanical subunit rotation. Therefore the $c_{10}$ ring of the $F_o$ part is driven by a proton concentration gradient across the membrane in combination with an electric potential, i.e. the 'proton motive force' [4]. Protons enter the first half-channel of subunit $a$ from the periplasm. They are transferred to an empty binding site on the $c_{10}$ ring . After stepwise rotation of the $c$ ring protons are released into the cytoplasm *via* the exit half-channel of subunit $a$ [5].

.................................................................................................................................................................................................

\* m.boersch@physik.uni-stuttgart.de; phone (49) 3641 933745; fax (49) 3641 933750; http://www.m-boersch.org

A central stalk, coupling the $F_1$ with the $F_O$ part, consists of subunits γ and ε. Due to an asymmetric structure of the γ subunit, the rotation of the central stalk causes large conformational changes in the $F_1$ part. The three αβ pairs of $F_1$ show a threefold symmetry. According to the concept of a 'binding-change-mechanism' [6] these subunits undergo a strict catalytic cycle and change their conformation sequentially in the order of 'open' to 'tight' to 'loose' in the case of ATP synthesis. Each conformation in this sequence shows a decreasing nucleotide binding affinity.

Over the last 15 years different experimental approaches were used to study the rotational behavior of the $F_oF_1$-ATP synthase [7-13]. First single molecule experiments studied the conformational changes of the $F_1$ part during ATP-hydrolysis *via* video microscopy. Therefore isolated $F_1$ parts were attached to a cover slip. In order to observe the stepwise rotation, a fluorescent actin filament was attached to the γ subunit. Data provided unequivocal evidence for a stepwise rotation of the γ subunit in discrete 120° steps during ATP hydrolysis [7]. Video dark field microscopy in combination with smaller fluorescent labels and a lower ATP concentration resolved two substeps of 80° and 40° [14, 15].

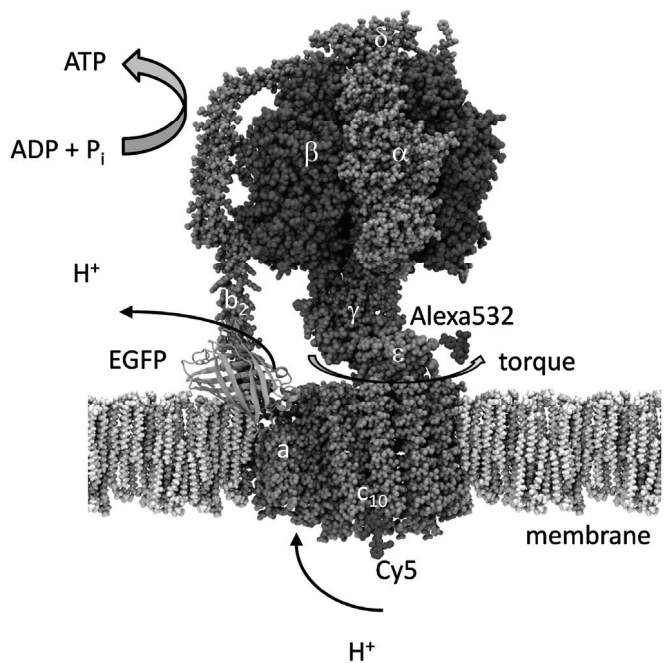

**Figure 1.** Model of the *E. coli* $F_oF_1$-ATP synthase embedded in the lipid membrane of a liposome. The $F_1$ motor consists of the non-rotating subunits $α_3β_3δ$ and the rotating γ and ε subunits (central stalk). The $F_o$ motor comprises the static subunits *a* and $b_2$ (peripheral stalk) and the rotating ring of ten *c* subunits.

In order to monitor subunit rotation in the holoenzyme $F_oF_1$-ATP synthase during ATP synthesis conditions, we used a different experimental approach based on Förster-type resonance energy transfer (FRET) [16-18]. Therefore we labeled the protein at different subunits with fluorescent markers [16, 19-24]. FRET is known as spectroscopic ruler in the range between 3 nm to 8 nm. The conformational changes associated with subunit rotation during the catalytic cycle of the protein are exactly in that range. Hence characteristic fluctuations in the intensity trajectory can be used to analyze the rotational steps. For the first single-molecule FRET experiments, the donor fluorophore was attached to a static *b*-subunit and the acceptor to the rotating γ-subunit of the protein [25, 26]. Analysis of single photon bursts showed FRET efficiency fluctuations and indicated three-stepped rotation of γ (and ε) at high driving forces as well as the opposite direction of rotation for ATP hydrolysis and synthesis [26-29]. To determine the step size of *c* ring rotation in the $F_o$ part the enzyme was labeled with EGFP at the static subunit *a* [30] and with Cy5 at one of the rotating *c* subunits. Using single molecule FRET the step size of $c_{10}$-ring rotation in $F_o$ was determined to 36° in ATP synthesis direction [31].

A recently developed improvement of this single-molecule approach uses three fluorescent markers (i.e. a triple-FRET experiment) to observe the coupled rotary motors of the $F_OF_1$-ATP synthase simultaneously [32, 33]. Therefore EGFP acted as static control at subunit *a*. Alexa532 was attached to the ε subunit and Cy5 to one of the *c* subunits. With a duty cycle-optimized alternating laser excitation scheme [32, 34, 35] (DCO-ALEX) including three different lasers, i.e. one for each fluorophore, it was possible to observe the conformational changes of the whole enzyme as well as the distance changes between subunits ε and *c*. The results of this single-molecule experiment indicate small and fast rotational steps between the major rotational steps of the ε subunit in 120° [33]. The custom-made software 'Burst Analyzer' [22] allows an automatic identification of photon bursts using intensity thresholds and to sort detected photons into different time windows according to their excitation laser pulse, which is a basic requirement for this kind of approach.

Therefore, the membrane protein was reconstituted into a artificial lipid bilayer or a lipid vesicle (liposome). The liposomes were allowed to diffuse freely through the confocal volume of about 10 fl. So every proteoliposome is generating a burst of photons. The large hydrodynamic radius of the vesicle allowed longer observation times. Thus it was also possible to mix buffers with different pH and $K^+$ concentrations to generate the appropriate proton motive force to drive ATP synthesis [36, 37]. Surface interactions of the protein can be minimized. No fluorescent filament is needed to observe the conformational changes which decisively affects the dwell times of the protein and the rates of ATP hydrolysis and ATP synthesis.

The two different step sizes of the $F_1$ motor and the $F_o$ motor indicate an intrinsic mismatch of the coupled rotary motors [38]. To accommodate this mismatch the protein has to store rotational energy within its subunits during the catalytic cycle. Energy transduction can be accomplished by internal elastic deformations of the central stalk. Although ongoing efforts have been made, the exact mechanism of energy transduction is not fully understood [39].

Associated with the first single-molecule experiments of $F_oF_1$-ATP synthase, different models of the energy transduction mechanism were published [40, 41]. One of the first models presumes that the free energy of nucleotide binding is converted into elastic strain energy with a mechanical efficiency of almost 100% [42]. Thereby the generated torque in the $F_1$ part was estimated to 45 pN nm. In this model the γ-subunit, consisting of intertwined helices, acts as an torsional spring between the $F_o$ and the $F_1$ part of the enzyme. This spring bridges the different step sizes of both rotary motors. Accordingly, the spring should be twisted up to 108° corresponding to three 36° steps of the $c_{10}$-ring [38]. After the loading process the elastic subunit releases its tension under liberation of one ATP molecule. Torque generation between different subunits of the $F_oF_1$-ATP synthase is still in the focus of scientific research, applying the fluctuation theorem for better estimations [43].

Using disulfide cross-linking experiments the stiffness of the internal and the protruding part of subunit γ was mapped [39, 44, 45]. Under the condition that both the internal and the protruding part of γ are compliant, a torsional angle of 42° was calculated at a torque of 40 pN nm. This angle is large enough to bridge both gears of the $F_oF_1$-ATP synthase [46]. Recently it was shown that the torsional energy is stored in elastic buffers on the central rotor and also in the lever of subunit β. Based on bending and twisting experiments of the subunit $b_2$ the authors addressed the stator of the $F_oF_1$-ATP synthase as an stiff scaffold for a strong connection of the $F_o$ and the $F_1$ part [45]. New molecular dynamics simulations predicted the amount of energy stored in each subunit of the central motor [47].

In this work we present a more detailed analysis of the conformational changes between the ε and *c* subunits during ATP hydrolysis using a single-molecule FRET approach. We aimed to unravel the symmetry mismatch by real time monitoring subunit rotation. Characteristic fluctuations of the intensity trajectories within single photon bursts indicated a reversible elastic deformations between the rotor parts of $F_o$ and $F_1$. With the knowledge of distance changes related to different conformational states it was possible to estimate the maximum angular displacement during the load-free rotation measurement.

# 2 EXPERIMENTAL PROCEDURES

## 2.1 $F_oF_1$-ATP synthase preparation for single-molecule FRET

In order to observe conformational changes during ATP hydrolysis, $F_oF_1$-ATP synthases from *Escherichia coli* were specifically labeled with two different fluorophores. We used genetically introduced reactive cysteines for labeling procedures with organic fluorophores. Alexa-532 was attached to the ε subunit at position 56, and Cy5 to one of the ten *c* subunits. Descriptions of plasmid constructions for the two *E. coli* $F_oF_1$-ATP synthase mutants as well as cell growth conditions, purification and labeling of the enzymes have been published previously [20, 21, 27, 33]. The double labeled $F_oF_1$-ATP synthase has been obtained as follows. At the *c*-subunit, the mutant included a reactive cysteine near the N-terminus as residue *c*2-cys, which was labeled with the FRET acceptor Cy5-monomaleimide (GE Healthcare). Labeling of the *c* ring with Cy5 was accomplished with detergent-solubilized $F_oF_1$-ATP synthase. Labeling efficiency was determined by UV-Vis absorption spectroscopy (using EGFP absorbance which was fused to the C-terminus of subunit *a* in this mutant as an internal protein concentration reference). The labeled enzymes were reconstituted to an excess of preformed liposomes, consisting of 90 % phosphatidylcholine and 10 % phosphatidic acid, with a mean diameter of about 150 nm [31].

Secondly, a different $F_oF_1$-ATP synthase mutant with a reactive cysteine in the ε subunit had to be used to introduce the FRET donor fluorophore Alexa532 specifically. The hydrophilic soluble $F_1$ part was prepared separately as described [27]. The rotating ε subunit in $F_1$ was labeled at residue position 56 with Alexa-532 maleimide (Invitrogen). Finally, the unlabeled $F_1$ part of the reconstituted double mutant was replaced by an Alexa-532-labeled $F_1$ part. $F_1$ parts were stripped off in a buffer without $Mg^{2+}$ and in the presence of EDTA. Alexa-532-labeled $F_1$ was rebound to $F_o$ in liposomes in the presence of $Mg^{2+}$ as published [26]. To control the replacement of $F_1$, ATP synthesis rates were measured after every preparation step. Aliquots of 2 μl of the reconstituted FRET-labeled $F_oF_1$-ATP synthases were shock-frozen in liquid nitrogen and stored at -80° C until use. For single-molecule FRET measurements during ATP hydrolysis an aliquot was thawed and diluted with 8 μl of liposome buffer. 0.5 μl of the diluted sample were mixed with 0.5 μl ATP (100 mM stock solution) and 49 μl of liposome buffer. The solution was pipetted immediately onto a glass cover slip. ATP hydrolysis was observed for 900 s in the custom-made confocal microscope (see below).

## 2.2 Confocal FRET microscope setup with DCO-ALEX

A custom-made inverted confocal microscope setup based on an Olympus IX 71 was used [48-60] for the FRET measurements with single $F_oF_1$-ATP synthases (Figure 2a). Two lasers were applied: a picosecond-pulsed laser at 532 nm (LDH-530, Picoquant) and a fiber-coupled picosecond-pulsed laser diode at 635 nm (LDH-P-635B, Picoquant, Berlin, Germany) with arbitrary repetition rates up to 80 MHz. Both laser sources were externally triggered and synchronized by an arbitrary waveform generator (AWG 2041, Tektronix) in order to get the required duty cycle-optimized alternating laser excitation scheme. The two laser beams were combined in order to provide a common confocal excitation volume (Figure 2b) using a dichroic beam splitter (DC 540, AHF Tübingen). To determine and adjust the exact position of each of the confocal laser spots in three dimensions, a sample of polychromatic polystyrene beads (transfluorospheres 488-635) embedded in a thin poly-vinyl-alcohol (PVA) film was used. After imaging an area of 5x5 $μm^2$ using a piezo-driven scan stage (P-517.2CD and PiFoc P-725.1CD, Physik Instrumente)[50, 54], the center of one polystyrene bead was located with an Gaussian fit algorithm to an accuracy of 20 nm with one laser. Subsequently the alignment of second laser for the two foci was improved.

To enter the confocal microscope the epi-fluorescence port was used. Beams were reflected by a dichroic filter (triple-band beam splitter 532/633, AHF Analysentechnik, Tübingen, Germany) and focused into the sample droplet with a water immersion objective (UPlanSApo60xW, 1.2 N.A., Olympus). The fluorescence signals passed a 150 μm pinhole and were split in two different detection channels using a beam splitter (DCXR 630, AHF) and were simultaneously detected by two avalanche photodiodes (SPCM-AQR-14, Perkin Elmer). In the first detection channel (APD1 in Figure 2A) the fluorescence of Alexa-532 was detected between 545 and 625 nm (HQ 585/80, AHF). A second detection channel (APD2) registered the signal of Cy5 between 663 nm and 737 nm (HQ 700/75, AHF). Single Photons were saved with two TCSPC cards (SPC 152, Becker & Hickl, Berlin, Germany) and a multi channel counter card for imaging (NI-PCI 6602, National Instruments) using our custom-made software.

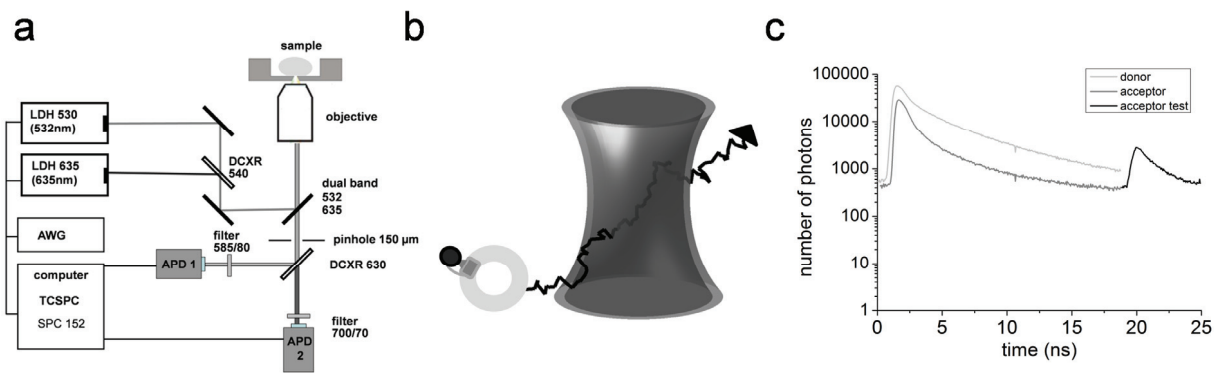

**Figure 2. (a)** Scheme of the confocal single-molecule microscopy setup used for FRET detection in solution. **(b)** Measurement principle of FRET in single proteoliposomes which are freely diffusing through the confocal volume. **(c)** Photon histogram of the laser excitation scheme. Two alternating laser pulses were applied in order to excite the FRET donor (532 nm) as well as the FRET acceptor directly (635 nm). Afterwards the recorded photons were sorted according to their arrival time. They were split into FRET donor photon (light grey curve, Alexa-532, between 0 ns and 18 ns in the microtime window), FRET acceptor photons (dark grey curve, Cy5, between 0 ns and 18 ns in the microtime window) and an acceptor test for Cy5 (black curve, between 18 ns to 25 ns in the microtime window, see text).

Both photon time traces were analyzed and combined by the software "Burst analyzer" (written by N. Zarrabi). With this tool it was possible to sort the photons into the correct time window according to the respective laser pulse. A first picosecond laser pulse at 532 nm (200 µW) was applied to excite the FRET donor fluorophore. The emitted photons of the FRET experiment, detected in both detection channels between 0 ns and 18 ns, were collected in a first time-window (Figure 2c). A second laser pulse at 635 nm (30 µW) was applied for the 'acceptor-test'. These photons were stored in the microtime window between 18 ns and 25 ns. With the help of this acceptor-test it was possible to sort out photophysical artifacts. The test was also used to analyze only dual-labeled enzyme and provided the intensity threshold criteria for automatic photon burst recognition.

## 3 RESULTS

To monitor the elastic deformations of single $F_oF_1$-ATP synthases we specifically attached two fluorophores to the enzyme from *E. coli*. Subunit ε of the $F_1$ part was labeled with Alexa-532 and one of the *c* ring subunits with Cy5. With this fluorophore configuration it was possible to observe the twisting of central stalk region during ATP hydrolysis. First single-molecule FRET measurements of the reconstituted enzyme revealed fluctuations in the FRET efficiency due to conformational changes. Figure 3 shows a selection of photon bursts from FRET-labeled $F_oF_1$-ATP synthases using intensity thresholds for automatic identification. In the lower panels, three different intensity trajectories are displayed. The light grey time trace indicates photons that belong to the direct excitation of the FRET donor fluorophore Alexa-532 by the 532-nm laser pulse. The dark grey timetrace shows the associated intensity trajectory of the FRET acceptor Cy5. In order to test whether the enzyme was labeled at the subunit c or not, a second laser pulse at the wavelength 635 nm was applied. The time trace of this laser pulse is displayed as a black line in the lower panels in Figure 3.

The upper panels of Figure 3 show the corresponding distance trajectories for each photon burst. For a maximum likelihood-based FRET distance trajectory we used the software "FRETtrace" (G. Schröder and H. Grubmüller [61]). This software was implemented into our FRET data analysis program "Burst Analyzer". Figure 3a shows a photon burst with only one large FRET efficiency change and smaller FRET fluctuations. After 60 ms the proteoliposome entered the confocal detection volume. The distance of the two fluorophores was about 5.8 nm. After 13 ms the distance of the two fluorophores decreased to only 4.5 nm. After that, the enzyme remained in this conformational state for about 25 ms, showing only minor fluctuations in the distance trajectory.

Figure 3b shows a photon burst with smaller alternating conformational changes. A series of distance changes between 4 and 4.5 nm could be recognized. In contrast, the photon burst of Figure 3c shows smaller as well as larger distance changes, indicating a highly active enzyme. The distances altered between 4 nm and 6 nm. The enzyme stayed for about 70 ms in the confocal spot, so that is a series of distance changes under ATP hydrolysis conditions were recorded. The last photon burst in Figure 3d shows probably a major, slow conformational change of the enzyme. The distance trajectory indicated a ramp-like back-and-forth movement. At first the distance increased from 4 nm to 5.9 nm. After that, the distance between the two fluorophores returned to its initial configuration.

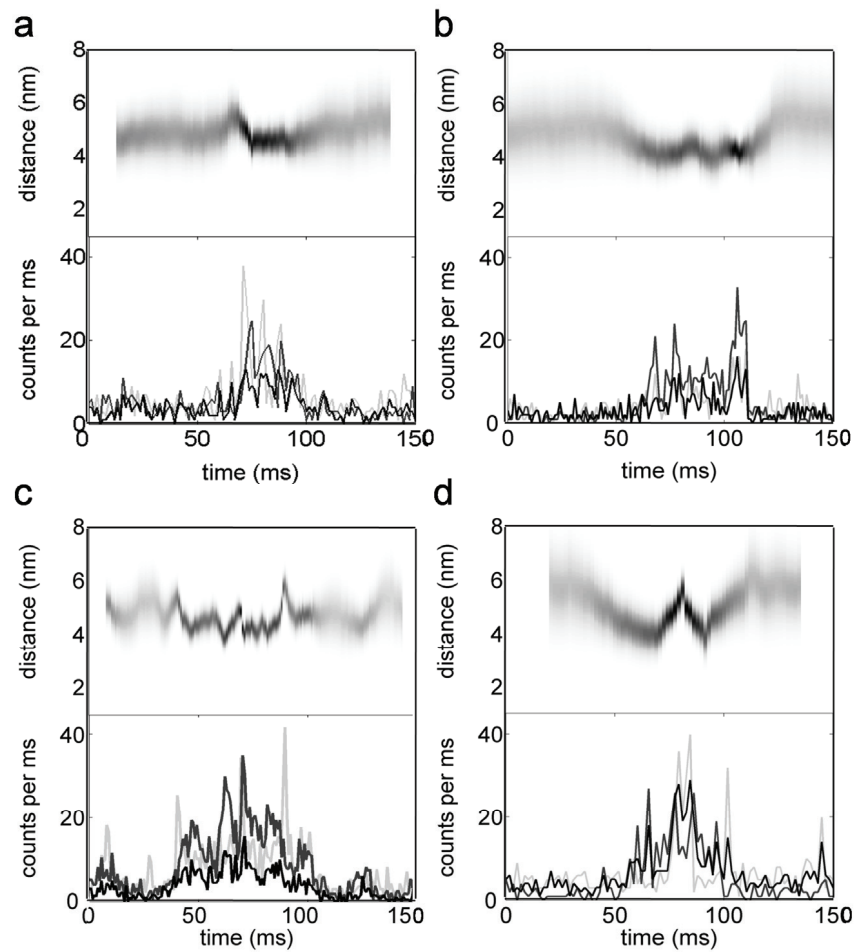

**Figure 3:** Photon bursts of single FRET-labeled $F_oF_1$-ATP synthases during ATP hydrolysis. Fluorescence intensity of the FRET donor Alexa-532 is shown as the light grey trace in the lower panels and intensity of the FRET acceptor Cy5 as the dark grey trace following 532 nm excitation. In order to test the presence of Cy5 at the $c$ ring a second alternating laser pulse was applied with 635 nm resulting in a third intensity trajectory (black lines). Fluctuations of the actual fluorophore distance, which corresponded to conformational changes within the enzyme, were calculated in the upper panels from the FRET efficiency and based on maximum likelihood estimations [61].

Now we analyzed the elastic deformation of the central stalk region of $F_oF_1$-ATP synthase quantitatively. Because single photon bursts showed only a parts of the whole catalytic cycle due to the limited observation time, we drew on statistical methods and compared the FRET levels of active enzymes with Monte Carlo simulations. With this approach it was possible to determine how far the rotor is deformed between the label positions at subunits ε and $c$ during catalysis. Because the stepping behavior of the $c$ ring takes place in 36° steps [21, 62], we decided to investigate how far the *c* ring rotates before one major conformational steps of the ε subunit follows. Briefly we simulated twisting angles of the rotor of 36°, 72°, 108° and 144°.

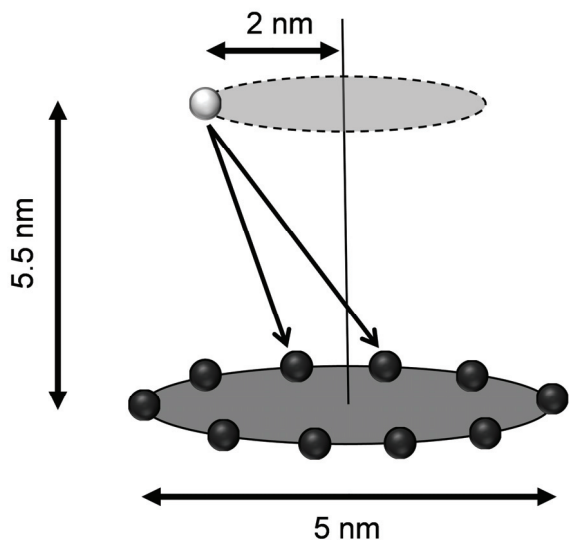

**Figure 4.** Geometric model of the relative distances changes between the two fluorophores attached to a single $F_oF_1$-ATP synthase (Alexa-532 on ε corresponds to the light grey dot; ten possible Cy5 positions on *c* correspond to the black dots) with respect to the axis of rotation. The distance parameters were used in Monte Carlo simulations of the elastic deformations during ATP hydrolysis.

Figure 4 shows the geometric model of the Monte Carlo simulations. We assumed a relative movement of the ε subunit (light grey sphere) with respect to *c* ring (black spheres). We used for the MC simulations the following parameters: (i) the distance between ε and *c* in z-direction was set to 5.5 nm, (ii) the position of the fluorophore on ε with respect to the axis of rotation was assumed to be 2 nm, and (iii) a diameter of the *c* ring was set to 5 nm. In order to figure out the influence of the brightness of the attached fluorophores, we also varied the apparent resolution of a position in our simulations. Given a Förster radius $R_0$=5.3 nm for Alexa-532/Cy5, 5000 FRET data points were simulated with mean sum intensities of 50 counts per ms (or 50 kHz, respectively), 150 count per ms (150 kHz) and 300 counts per ms (300 kHz). For each sum intensity FRET transition density plots for twisting angles of 36°, 72°, 108° and 144° were calculated. We assumed a FRET distance error of 0.5 nm for each distance, so that the FRET transition distributions were blurred.

The FRET transition density plot connects the distance of the two fluorophores before and after a conformational change [63]. A pair of distance values appears as one data point in the density plot. Especially for small intensities, the simulated FRET transition density plots were not distinctive enough. The distance pair distribution was blurred, so that it was difficult to distinguish between the different twisting angles. Therefore we plotted the FRET distance 1 versus the FRET distance change $\Delta d_{1,2}$ in order to find more pronounced differences. In the experimental FRET data it was not possible to assign small distance changes less than 0.5 nm. Therefore we applied a third FRET change density plot. By plotting the simulated pairs of subsequent FRET distance changes of $\Delta d_{1,2}$ versus $\Delta d_{2,3}$ the different twisting angles was easier to discriminate. Taking all three plots together it seemed possible to determine the maximum twist angle of the rotor in $F_OF_1$-ATP synthases during ATP hydrolysis from the comparison of the simulated FRET transition plots with the measured single-molecule FRET data.

The results of the MC simulation with a mean sum intensity of 50 kHz is shown in Figure 5. All FRET transition plots showed broad distributions independent of the twisting angle. No preferred accumulation points of the different stop positions of the associated conformational changes were found. The simulation with a low mean sum intensity revealed that the fluorophores need high quantum efficiencies as well as optimized excitation wavelengths and duty cycle-optimized laser excitation schemes to determine elastic deformations of the central stalk from the recorded single-molecule FRET experiments. Otherwise the measured FRET distributions were too broad to distinguish between the single conformational states of the rotor during the catalytic cycle.

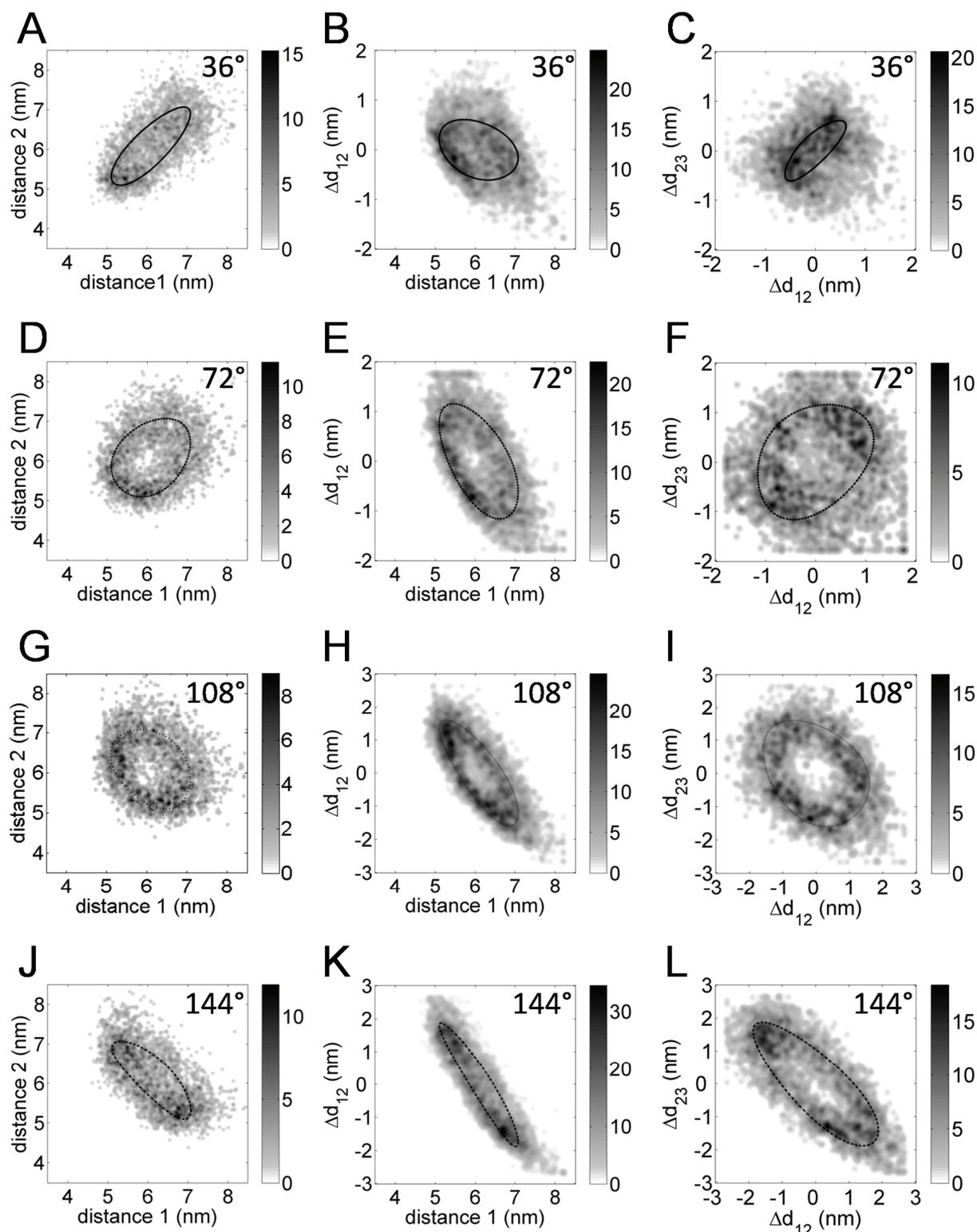

**Figure 5.** Monte Carlo simulations of the distance changes in the rotor of $F_oF_1$-ATP synthases via FRET for relative movements of *c*-subunits *versus* ε. The mean fluorescence intensity for this simulation was set to 50 counts per ms. The rows show the transition plots for twisting angles between ε56 an *c*2 of 36° (**A, B, C**), 72° (**D, E, F**), 108° (**G, H, I**) and 144° (**J, K, L**).

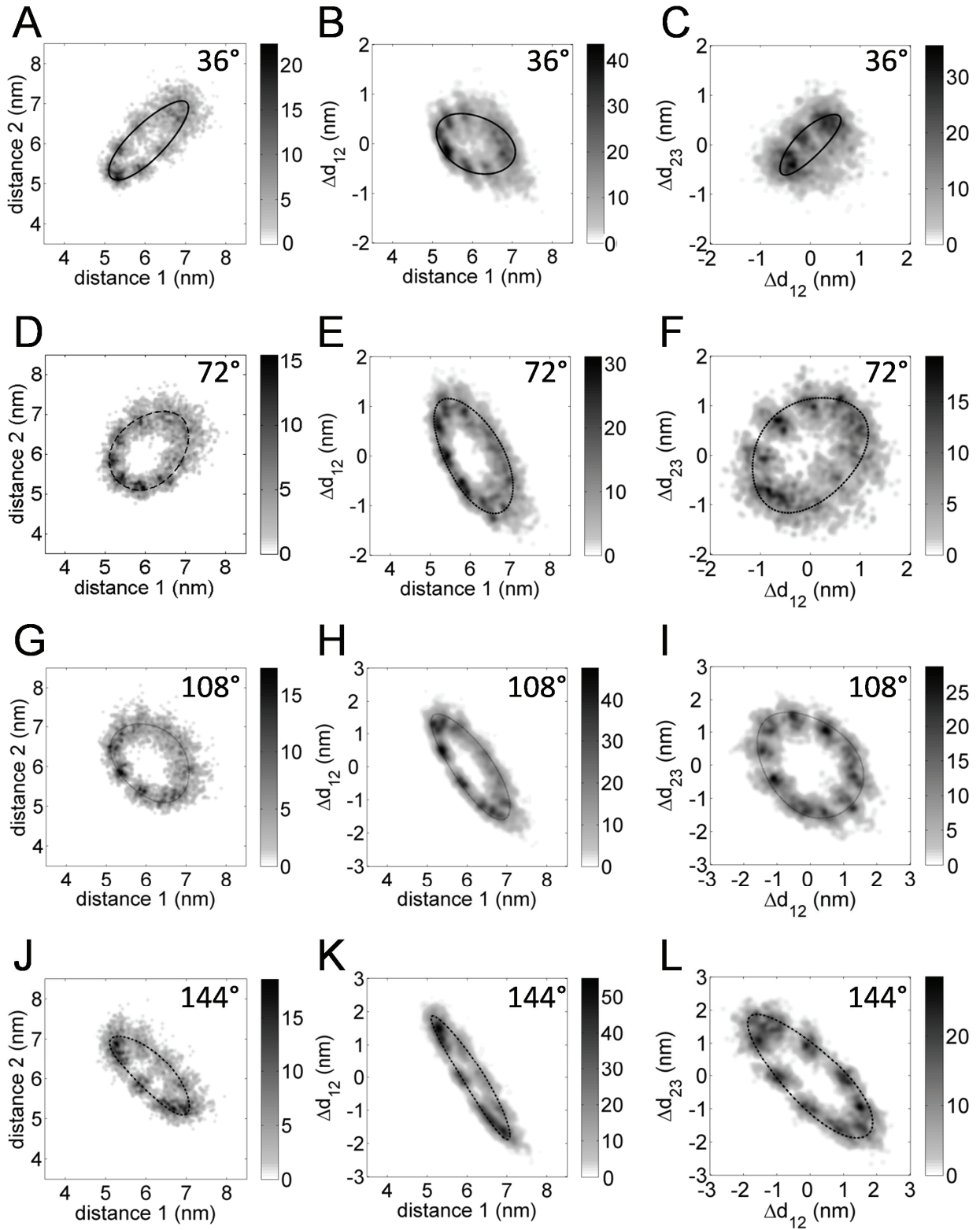

**Figure 6.** Monte Carlo simulations of the distance changes in the rotor of $F_oF_1$-ATP synthases via FRET for relative movements of $c$-subunits *versus* ε. The mean fluorescence intensity for this simulation was set to 150 counts per ms. The rows show the transition plots for twisting angles between ε56 an $c$2 of 36° (**A, B, C**), 72° (**D, E, F**), 108° (**G, H, I**) and 144° (**J, K, L**).

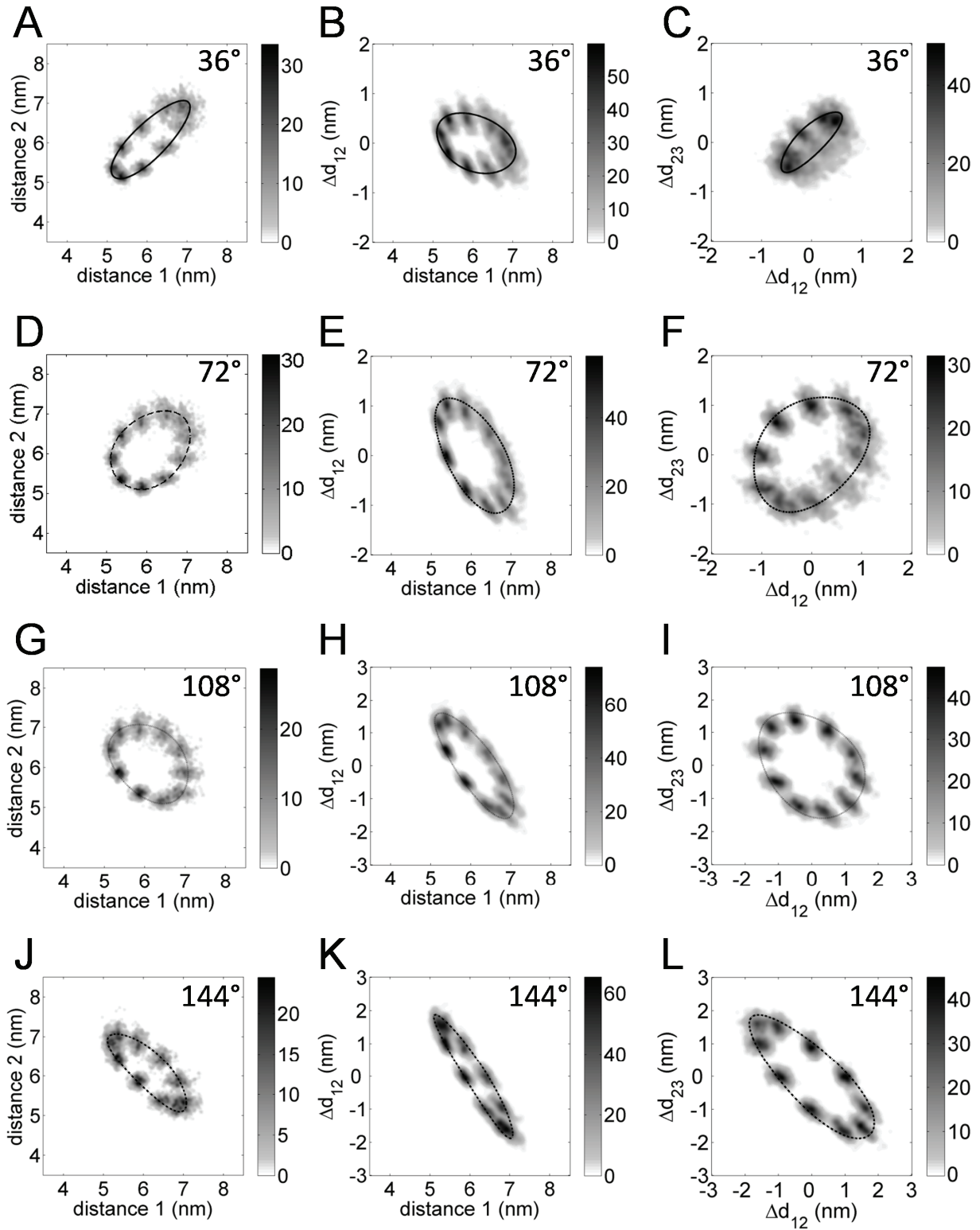

**Figure 7.** Monte Carlo simulations of the distance changes in the rotor of $F_oF_1$-ATP synthases via FRET for relative movements of *c*-subunits *versus* ε. The mean fluorescence intensity for this simulation was set to 300 counts per ms. The rows show the transition plots for twisting angles between ε56 an *c*2 of 36° (**A, B, C**), 72° (**D, E, F**), 108° (**G, H, I**) and 144° (**J, K, L**).

The simulation with a mean sum intensity of 150 kHz (Figure 6) revealed a better distance resolution. The distributions were much narrower and showed distinct accumulation points representing single conformational changes. The transitions for the different twisting angles were now clearly distinguishable and in good agreement with the theoretical ellipses. The maximum distance changes of two consecutive conformational changes had decreased. They reached from 0.8 nm for the 36° distribution up to 2.6 nm for the 144° distribution. The FRET transition density plot (Figure 6A) also showed, that it was still difficult so distinguish single transition steps even with an increased sum intensity. A 36° twist caused only small FRET efficiency changes and those will be difficult to assign manually in experimental data.

Figure 7 shows the different transitions plots based on simulations with a mean sum intensity of 300 kHz. Compared to the density plots discussed before, the resolution of the single stopping positions improved slightly. The absolute distances as well as the distance changes were in good agreement with the theoretical determined values. For these theoretical fluorescence intensities, one has to take into account, that this minimal improvement of the resolution could dramatically enhance photobleaching. The goal for this measurements is to maximize the excitation intensity without bleaching the attached fluorophores. Considering the distance distributions of all three density plots, the simulated values seemed promising to further investigate the elastic deformations of the $F_OF_1$-ATP synthase. But they also showed, that large single-molecule FRET data sets have to be used to determine the twisting angle between the subunits ε and $c$.

## 4  DISCUSSION

In order to address the question about the localization of elastic energy storage during the catalytic cycle of the $F_oF_1$-ATP synthase, we presented a new tool to analyze the twisting angle of the internal rotor subunits based on MC simulation. Single-molecule as well as other techniques have unraveled, that the rotor subunits in the hydrophilic $F_1$ part of the enzyme rotate in discrete 120° steps [7, 27]. On the other hand the $c$-ring of the membrane bound $F_o$ part showed single 36° steps [31]. For ATP hydrolysis and ATP synthesis, an efficient energy transduction mechanism is needed to bridge the two different step sizes.

Conformational changes of the $F_oF_1$-ATP synthase were visualized using two different fluorophores attached to the ε subunit (Alexa532) and to one of the $c$ subunits of the $c$ ring (Cy5). Single-molecule FRET measurements revealed distance changes of the fluorophores with sub-nanometer resolution. Single enzymes were reconstituted into liposomes and diffused freely through the confocal volume. Photon burst were identified automatically from the time traces based on intensity threshold criteria. A DCO-ALEX laser excitation scheme allowed to remove photophysical artifacts as well as incompletely labeled proteins.

First ATP hydrolysis measurements indicated elastic deformations of the internal stalk during the catalytic cycle (Figure 3). FRET level changes were manually assigned within the photon bursts. Previous elasticity measurements of the rotor subunits using surface-attached $F_1$ sectors and with $F_oF_1$-ATP synthases in detergent supported an elastic behavior [44, 46]. Twisting angles of more than 40 degrees were found within the γ/ε and $c$-subunits by observing the movement of large beads on the enzyme. On the other hand the same experimental approaches have shown, that the peripheral stalk, consisting of the right handed coiled-coil structure of the dimeric $b$ subunits, is rather stiff. Here the twisting angles of the subunits were about 10° [45].

To determine the resolution limits required to identify the twisting angle of the central stalk *via* single-molecule FRET measurements, we simulated the rotation of the $c$ ring compared to a fixed fluorophore position at the subunit ε, so that FRET distributions of the simulated data could be compared with the measurement. We decided to simulate rotor twisting in steps of 36° as well as multiple steps of the $c$ ring omitting some of the stopping positions. The rotation of the $c$ ring were then 72°, 108° and 144°. We determined the distances of the two fluorophores before and after a conformational change and plotted the results as a 2D density plots. Furthermore, distance changes of two consecutive rotational steps and the absolute FRET-distance 1 versus the distance change were applied to distinguish the twisting angle more clearly. The influence of the mean sum brightness of the fluorophores with respect to the broadening of the distance distribution was evaluated. The simulated data showed, that a low mean sum intensity corresponded to a broad distribution of the single FRET data pairs. No clear stopping positions were recognizable. This made it difficult to distinguish between the single conformational changes, and, therefore, to estimate the twisting angle of the rotating subunits. By increasing the mean sum intensity to 150 kHz, the distance distributions were clearly defined, and further increase did not improve the resolution significantly. As high laser excitation would cause photobleaching of the attached fluorophores, the observation time of a single FRET-labeled $F_OF_1$-ATP synthase will decrease.

Single-molecule FRET measurements are excellent tools to study the energy transduction mechanism within the $F_OF_1$-ATP synthase. With this approach it is possible to investigate transient elastic deformations of the rotor subunits as well as their twisting angle. In principle the whole enzyme can be mapped by this method in order to determine the stiffness of different subunits under catalytic turnover conditions. To determine the twisting angle of the rotor precisely, good statistical data have to be obtained by future single-molecule FRET experiments. There is also the need of new fluorophores with a high quantum yield that lead to well defined distance distributions with a minimal excitation intensity to avoid photobleaching [64]. Most importantly, the observation time has to be prolonged in order to detect the sequential conformational changes of a single enzyme while it stays in the confocal volume. One possible solution is the electrokinetic trap ('ABELtrap') developed by A E Cohen and W E Moerner [65-67]. In preliminary experiments, trapping a FRET-labeled $F_oF_1$-ATP synthase in the laser focus and hold it there until the fluorophores photobleached led to observation times up to 10 seconds.

**Acknowledgements**

This work was in part supported by the DFG grants BO 1891/10-1 and BO 1891/10-2 to M.B.. The authors want to thank Prof. Dr. S. D. Dunn (Western University of Ontario, London, Canada) and Prof. Dr. P. Gräber for their ongoing support in $F_oF_1$ mutant constructions and enzyme preparation, and Prof. Dr. J. Wrachtrup for help with the microscope.